\documentstyle[aps]{revtex}
\newcommand{\be}{\begin{equation}\label}
\newcommand{\ee}{\end{equation}}
\newcommand{\bea}{\begin{eqnarray}\label}
\newcommand{\eea}{\end{eqnarray}}
\newcommand{\ul}{\underline}
\newcommand{\pr}{\partial}
\newcommand{\eps}{\varepsilon}
\renewcommand{\Re}{\mbox{Re}}
\renewcommand{\Im}{\mbox{Im}}
\newcommand{\vPhi}{\ul{\Phi}}
\newcommand{\vpsi}{\ul{\psi}}
\newcommand{\vu}{\ul{u}}
\newcommand{\vv}{\ul{v}}
\newcommand{\vhu}{\ul{\hat{u}}}
\newcommand{\vhv}{\ul{\hat{v}}}
\newcommand{\mL}{\ul{\ul{L}}}
\newcommand{\mU}{\ul{\ul{U}}}
\newcommand{\mM}{\ul{\ul{M}}}
\newcommand{\id}{\ul{\ul{1}}}
\newcommand{\Ga}{\ul{\Gamma}_a}
\newcommand{\Gb}{\ul{\Gamma}_b}
\newcommand{\Gc}{\ul{\Gamma}_c}
\newcommand{\tG}{\ul{\Gamma}}
\newcommand{\odr}{{\cal O}}
\newcommand{\mR}{\ul{\ul{R}}}
\newcommand{\tvPhi}{\tilde{\vPhi}}
\newcommand{\tL}{\tilde{\cal L}}
\newcommand{\tN}{\tilde{\cal N}}
\newcommand{\tvpsi}{\tilde{\vpsi}}
\newcommand{\tGa}{\ul{\tilde{\Gamma}}_a}
\newcommand{\tGb}{\ul{\tilde{\Gamma}}_b}
\newcommand{\ttG}{\tilde{\tG}}
\newcommand{\tvu}{\tilde{\vu}}
\newcommand{\BbbR}{I\!\! R}
\newcommand{\BbbC}{I\!\!\! C}
\begin{document}
\title{Weakly nonlinear analysis in
spatially extended systems as a formal perturbation scheme}
\author{Wolfram Just\thanks{e--mail: 
wolfram@arnold.fkp.physik.th-darmstadt.de}}
\address{Max--Planck--Institut f\"ur Physik komplexer Systeme,
Bayreuther Stra\ss e 40, Haus 16, D--01187 Dresden, Germany}
\author{Frank Matth\"aus, Hans Rainer V\"olger,  Christine Just,
Benno Rumpf, and Anja Riegert} 
\address{Theoretische Festk\"orperphysik,
Technische Hochschule Darmstadt,
Hochschulstra\ss e 8, D--64289 Darmstadt, Germany}
\date{June 1, 1997}
\maketitle
\begin{abstract}
The well known concept, to reduce the spatio--temporal dynamics
beyond instabilities of trivial states to amplitude modulated patterns,
is reviewed from the point of view of a formal perturbation expansion for
general dissipative partial differential equations. For codimension one
instabilities closed analytical formulas for all coefficients of the
resulting amplitude equation are given, with no further restriction on the
basic equations of motion. Both the 
autonomous and the explicitly time--dependent
case are discussed. For the latter, the problem of strong resonances is 
addressed separately. The formal character of the expansion allows for
an analysis of higher--codimension instabilities like the Turing--Hopf
instability and for the discussion of principal limits of the amplitude 
approach in the present form.
\end{abstract}
\section{Introduction}
Pattern formation in dissipative systems under
non--equilibrium conditions is a classical field of physical science and has
in particular developed from hydrodynamic problems. 
In addition, this subject has become recently very popular 
in the context of optical, chemical, magnetic, and even biological systems 
(cf.~ref.\cite{CroHo} and references theirin). The renewed interest was 
partially stimulated by developments in nonlinear dynamics.

Theoretical approaches rely strongly on analytical perturbation expansions,
in order to reduce the equations of motion to dynamical relevant quantities. 
To some extent these concepts are limited to a neighbourhood of an
instability of a simple state. Such concepts
have proven to be powerful tools in the context of low--dimensional
dynamical systems, even with quite mathematical rigour (e.g.~\cite{GuHo}).
For spatially extended systems with a large number of relevant degrees 
of freedom (i.e.~the limit of 
large aspect ratio in the hydrodynamic context) 
such approaches have been introduced based on a multiple scale 
analysis \cite{NewW,Seag}. They have been applied to a huge number of
concrete examples and can even be found in textbooks (e.g.~\cite{Mann}).
In fact the corresponding reduced description beyond simple instabilities, 
the Ginzburg--Landau equation, is very well known, although
its properties are extremely complex and are a subject of intensive
research. 

Like in normal forms for differential equations, the details of the
equations of motion, so to say the ''physics'', is to a large extent
contained in the coefficients of the reduced equation. Hence it is
desirable to have a closed expression at hand, to determine these
quantities for an arbitrary equation of motion.
Such formulas can be partially found in the literature
\cite{Kura76,Haken,Elmer89}, but unfortunately certain restrictions on the
structure of the equation or the instability are imposed, which may limit
the applicability of such expressions. It is one goal
to fill this gap and to give these formulas for a large class
of equations of motion { including the case of explicitly time dependent
systems.} Hence for their evaluation,
which only requires some simple algebraic computations,
it is not necessary to perform the multiple scales analysis in each
concrete case. { In this sense our results apply easily to almost all
evolution equations discussed in the physical context,
considerably simplify the explicit computation of an amplitude equation,
and show the common algebraic structure among the explicit expression for the
coefficients as far as they are available in the literature.

In order to keep the presentation self--contained 
we review the complete derivation of amplitude equations
within the well known concept of multiple scales analysis. 
Although our formulation follows the standard lines,
we stress the following properties.
The expansion can be understood as a formally exact
procedure, and no physical assumptions (e.g.~on scales), 
sometimes used to simplify calculations, are necessary. In addition, 
our approach respects the vector type structure inherent in the underlying
equation of motion, and the final expressions can be understood as scalar
quantities with respect to this structure. Hence with a suitable
notation the full calculation is by no means more involved as for simple
model equations. This observation has two
important consequences. On the one hand it opens the possibility to analyse
higher--codimension instabilities starting from physical equations of 
motion, by going to higher orders in the multiple scales analysis. 
On the other hand one can discuss the principal limits of the multiple scales
approach in extended systems. In fact we will present one such
example, which occurs within the class of explicitly time dependent systems.
Finally our approach clearly distinguishes} between 
the multiple scales expansion and separate approximations for the evaluation 
of the coefficients, which may be mixed if the concept is applied
to a definite equation of motion.

For those readers who are not interested in technical details, section 
\ref{sec2} summarises the notation and the results for the instability of a
single mode, i.e.~the explicit formulas for the coefficients of the 
Ginzburg--Landau equation. A brief discussion of the result is 
supplemented in section \ref{sec4}. The restriction to one spatial dimension
is far from being only technical, since a formally satisfactory and
sufficient general approach for spatially higher--dimensional 
rotationally symmetric systems is still
missing, if one disregards the Newell--Whitehead--Segel treatment of
nearly one--dimensional extended patterns.
The technical derivation is reviewed in section \ref{sec3} for autonomous
systems. As an application to degenerated instabilities, the analysis for
the Turing--Hopf instability together with the explicit formulas for
the complete set of coupled amplitude equations is presented in section
\ref{sec7}. The peculiarities which arise in the presence of explicitly
time--dependent equations are analysed in section \ref{sec5}. 
The phenomenon of strong resonances, which is associated with this
situation, is discussed in section \ref{sec6}. Finally
a few remarks on other higher--codimension instabilities and the 
generalisations for the discussion of spatially
non--homogeneous situations are given. 
\section{Basic Notation and Result}\label{sec2}
We consider a physical system being invariant with respect to translations
in space
as well as in time. In order to deal with very general
situations we allow for an $N$--component real field $\vPhi(x,t)$. 
A trivial, that means spatially homogeneous and time--independent, 
state should undergo an instability. Without loss of generality we
may assume that its value is zero, so that the evolution equation reads
\be{ba}
\frac{\pr \vPhi}{\pr t} = {\cal L} \vPhi + {\cal N}[\vPhi],
\qquad \vPhi \in {\BbbR}^N,\quad x \in {\BbbR} \quad .
\ee
Here the linear operator ${\cal L}$ governs the instability and
${\cal N}$ denotes all the nonlinear contributions 
(cf.~eqs.(\ref{bha}) and (\ref{bhc})).

We presuppose that on variation of the system parameters one mode
with wavenumber $q_c$ and frequency $\omega_c$ becomes 
unstable\footnote{Either the wavenumber or the frequency may vanish.}.
If we measure the deviation from this instability by the small
quantity $\eps^2$, then the linear operator can be cast into
\be{bb}
{\cal L} = {\cal L}^{(0)} + \eps^2 {\cal L}^{(2)} + \odr(\eps^4) \quad.
\ee
The most general expression for the linear operator at the threshold,
which is compatible with the space--time translation invariance,
reads
\be{bc}
{\cal L}^{(0)} \vpsi = \sum_{\alpha} \mL_{\alpha}
\frac{\pr^{\alpha}}{\pr x^{\alpha}} \vpsi \quad,
\ee
where the real matrices $\mL_{\alpha}$ take the vector character of the
field into account\footnote{ 
The formal expression (\ref{bc}) for the linear 
operator was written down for convenience only, since a lot of
evolution equations contain derivatives of finite order. 
Nevertheless the case of an infinite 
series, i.e.~integral operators, is permitted too. It is evident from
eqs.(\ref{bda}), (\ref{bdc}) and the subsequent derivation, that our
approach applies, if the kernel and the moments up to order three admit
a Fourier transform.}.
A similar expression can be also written down for the second order 
contribution, but it is of no special
use in the sequel. The (right--)eigenvalue
problem of the operator (\ref{bc}) can be solved in terms of Fourier--modes
\bea{bd}
{\cal L}^{(0)} e^{i k x} \vu_k^{(\nu)} 
&=& \lambda^{(\nu)}(k) e^{i k x}  \vu_k^{(\nu)} 
\label{bda}\\
\mL(k) \vu_k^{(\nu)} &=&
\lambda^{(\nu)}(k) \vu_k^{(\nu)}\label{bdc}\\
\mL(k) &:=&  \sum_{\alpha}  (i k)^{\alpha} 
\mL_{\alpha}\label{bdd} \quad .
\eea
Here the index $\nu$ numbers the different branches of the eigenvalue 
problem. The eigenvalues as well as the
vector part $\vu_k^{(\nu)}$ of the eigenfunctions 
are completely determined by the complex
$N$--dimensional algebraic equation (\ref{bdc}).
By presupposition all eigenvalues have a negative real part except for
one branch $\nu=\nu_c$, where the real part vanishes 
at $k=q_c$\footnote{For $q_c=0$ a pair of complex conjugate eigenvalues
occurs, because of symmetry. Then $\nu_c$ denotes one of these branches.}
\be{be}
\Re \lambda^{(\nu_c)}(k)<0 \,\, (|k|\neq |q_c|) ,\quad
\lambda^{(\nu_c)}(q_c) =: i \omega_c,\quad \vu_{q_c}^{(\nu_c)} =: \vu_c
\quad .
\ee
Finally, we need for technical purposes the left--eigenvectors
of the matrix (\ref{bdd}) 
\be{bf}
\vv_k^{(\nu) *} \mL(k) = \vv_k^{(\nu) *} \lambda^{(\nu)}(k) \quad ,
\ee
and especially the eigenvector at the threshold $\vv_c:=\vv_{q_c}^{(\nu_c)}$.

Let us now turn to the nonlinear contributions. Of course they depend on the
system parameters, that means on $\eps^2$, too. But to the perturbation
expansion only the expression at the threshold $\eps=0$ and 
terms of second or third order in the field amplitude contribute.  
Hence if we put
\be{bg}
{\cal N}[\vpsi] = {\cal N}_2[\vpsi]  + {\cal N}_3[\vpsi]  
+ \odr(\eps^2,\|\vpsi\|^4) \quad ,
\ee
then the most general expressions of second and third order read
\bea{bh}
{\cal N}_2[\vpsi] &=& \sum_{\alpha \beta} 
C^{(\alpha \beta)}\left\{
\frac{\pr^{\alpha} \vpsi}{\pr x^{\alpha}},
\frac{\pr^{\beta} \vpsi}{\pr x^{\beta}}\right\} \label{bha}\\
\left(C^{(\alpha \beta)} \left\{\vu,\vv \right\}\right)_j &:=&
\sum_{l m} c_{jlm}^{(\alpha \beta)} u_l v_m \label{bhb}\\
{\cal N}_3[\vpsi] &=& \sum_{\alpha \beta \gamma} 
D^{(\alpha \beta \gamma)}\left\{
\frac{\pr^{\alpha} \vpsi}{\pr x^{\alpha}},
\frac{\pr^{\beta} \vpsi}{\pr x^{\beta}},
\frac{\pr^{\gamma} \vpsi}{\pr x^{\gamma}} \right\} \label{bhc}\\
\left(D^{(\alpha \beta \gamma)} \left\{\vu,\vv, \ul{w} \right\}\right)_j 
&:=& \sum_{l m n} d_{jlmn}^{(\alpha \beta \gamma)} u_l v_m w_n 
\label{bhd} \quad.
\eea
Here the real tensors (\ref{bhb}) and 
(\ref{bhd}) take the vector character of the equation into 
account\footnote{In the sequel the symmetry relations
\begin{displaymath}
C^{(\alpha\beta)}\{\vu,\vv \}=C^{(\beta \alpha)}\{ \vv,\vu \}
\quad 
D^{(\alpha \beta \gamma)}\{\vu,\vv,\ul{w} \}=
D^{(\beta \alpha \gamma)}\{\vv,\vu,\ul{w} \}=\ldots
\end{displaymath}
are used.}.

In order to investigate the motion beyond the instability, we expand the
solution of eq.(\ref{ba}) in terms of the small parameter $\eps$, by taking
explicitly the amplitude modulation of the marginally stable mode into 
account
\be{bi}
\vPhi(x,t) = \eps \left[ \vu_c e^{i q_c x+ i \omega_c t} 
A(\tau_1,\tau_2,\ldots, \xi_1,\xi_2,\ldots) + 
\vu_c^* e^{-i q_c x- i\omega_c t} A^*(\ldots) \right] 
+ \eps^2 \vPhi^{(2)} + \eps^3 \vPhi^{(3)}+ \odr(\eps^4) \quad .
\ee
Here the abbreviations
\be{bj}
\tau_n := \eps^n t,\quad \xi_n := \eps^n x
\ee
determine the scales of the slowly varying amplitude 
$A$. The evolution equation for $A$ can be derived
if we require that this representation 
does not contain secular
contributions (cf. section \ref{sec3}). 
This procedure results in the well known
Ginzburg--Landau equation
\be{bk}
\left(\frac{\pr }{\pr \tau_2} -v \frac{\pr}{\pr \xi_2}\right) A
= \eta A + r |A|^2 A +D \frac{\pr A}{\pr \xi_1^2} \quad .
\ee
Here the convective velocity $v$ and the
diffusion coefficient $D$ are given in terms of derivatives of the
critical eigenvalue
\bea{bl}
v &=& \left. \Im \frac{d \lambda^{(\nu_c)}(k)}{d k} \right|_{k=q_c}  
\label{bla}\\
D &=& -\frac{1}{2} \left. \frac{ d^2 \lambda^{(\nu_c)}(k)}
{d k^2}\right|_{k=q_c}\label{blb}  \quad ,
\eea
the coefficient of the cubic term is determined by the nonlinearities
\bea{bm}
r &=& \frac{ ( \ul{v}_c | \tG+\ul{\Delta}) }
{ (\ul{v}_c | \ul{u}_c ) }\label{bma} \\
\tG
&=& 2 \sum_{\alpha} (i q_c)^{\alpha} C^{(\alpha 0)}\left\{\vu_c,
2 \Ga \right\} 
+ 2 \sum_{\alpha\beta} (2 i q_c)^{\beta} (- i q_c)^{\alpha}
C^{(\alpha \beta)}\left\{\vu_c^*,
\Gb \right\} \label{bmb} \\
\Ga &:=&  - \frac{1}{\mL(0)}
\sum_{\alpha \beta} (i q_c)^{\alpha} (-i q_c)^{\beta}
C^{(\alpha \beta)}\left\{\vu_c,\vu_c^*\right\} 
\label{bmc}\\
\Gb &:=& - \frac{1}{
\mL(2 q_c) - 2 i \omega_c \id}
\sum_{\alpha \beta} (i q_c)^{\alpha} (i q_c)^{\beta}
C^{(\alpha \beta)}\left\{\vu_c,\vu_c\right\} \label{bmd} \\
\ul{\Delta} &=& 3 \sum_{\alpha\beta\gamma}
(iq_c)^{\alpha} (iq_c)^{\beta} (-iq_c)^{\gamma}
D^{(\alpha\beta\gamma)} \left\{\vu_c,\vu_c,\vu_c^*\right\}\label{bme}
\quad ,
\eea
and the coefficient of the linear term is given by the matrix
element of the perturbation ${\cal L}^{(2)}$
\be{bn}
\eta = \frac{ q_c/(2\pi) \int_0^{2\pi/q_c} \left( \vv_c e^{i q_c x}
| {\cal L}^{(2)} | \vu_c e^{i q_c x}\right) \, dx} 
{ (\ul{v}_c | \ul{u}_c )} \quad .
\ee
The integral\footnote{The case $q_c=0$, in which only the scalar products
appear, is captured by this notations as the formal limit
$q_c\rightarrow 0$.}
just picks out the Fourier component with wave number
$q_c$. $(.|.)$ denotes the usual 
scalar product in ${\BbbC}^N$.
For convenience appendix \ref{appc} contains the evaluation of these
expressions for the Maxwell--Bloch equations as an example.
\section{Derivation of the Amplitude Equation}\label{sec3}
One inserts the expression (\ref{bi}) into eq.(\ref{ba}) and expands
order by order taking the notation (\ref{bb}), (\ref{bc}), (\ref{bg}),
(\ref{bha}),
and (\ref{bhc}) into account.

To first order in $\eps$ only the linear operator contributes. Since
the amplitude $A$ acts as a constant at this order one obtains
\be{ca}
i \omega_c \vu_c e^{i q_c x +i\omega_c t} A -
i \omega_c \vu_c^*  e^{-i q_c x -i\omega_c t} A^* 
=\mL(q_c) \vu_c e^{i q_c x+ i\omega_c t} A 
+\mL(-q_c) \vu_c^* 
e^{-i q_c x -i \omega_c t} A^* \quad .
\ee
By virtue of the relation $\mL(k)^*=\mL(-k)$ this expression is nothing else
but the eigenvalue equation
(\ref{bdc}) of the critical mode (cf.~(\ref{be})).

To second order in $\eps$ one gets
\bea{cb}
& &\frac{\pr \vPhi^{(2)}}{\pr t} + \vu_c e^{i q_c x + i \omega_c t} 
\frac{\pr A}{\pr \tau_1} + \vu_c^* e^{-i q_c x- i \omega_c t} 
\frac{\pr A^*}{\pr \tau_1} \\
&=& \left({\cal L}^{(0)} \vPhi^{(2)}\right)^{[0]} +  
\left({\cal L}^{(0)} \vu_c e^{i q_c x + i \omega_c t} A\right)^{[1]} +
\left({\cal L}^{(0)} \vu_c^* e^{-iq_c x- i \omega_c t} 
A^*\right)^{[1]} + \left( {\cal N}_2[\vPhi]\right)^{[2]} \quad . \nonumber
\eea
Here the additional superscript characters 
indicate at which order in $\eps$ the expression 
in the bracket has to be evaluated. 

First let us consider the linear terms. From the identity
\be{ce}
\left( \frac{\pr^{\alpha}}{\pr x^{\alpha}} e^{i q_c x} A\right)^{[1]} 
= \alpha (i q_c)^{\alpha-1 } 
e^{i q_c x} \frac{\pr A}{\pr \xi_1} 
\ee
and the definition (\ref{bc}), the relation
\be{cf}
\left( {\cal L}^{(0)} \vu_c e^{i q_c x + i \omega_c t} A\right)^{[1]} =
-i \mL'(q_c) \vu_c e^{i q_c x + i \omega_c t} 
\frac{\pr A}{\pr \xi_1}
\ee
follows.
In general $\vu_k^{(\nu_c)}$ depends on the wavenumber, so that $\vu_c$
is not an eigenvector of the derivative $\mL'(q_c)$. However, as shown in
appendix \ref{appA}, we can consider this special case without imposing any 
restriction on the validity of our results. Although this step is by no
means essential, it helps to simplify considerably the following
computations. By taking the derivative of the eigenvalue equation 
(\ref{bdc}) with respect to $k$ 
at $k=q_c$ and recalling that the real part has a maximum at $q_c$, one 
obtains
\be{cfa}
\mL'(q_c) \vu_c = i v \vu_c
\ee 
taking the definition (\ref{bla}) into account. Hence eq.(\ref{cb}) 
simplifies to
\be{cg}
\frac{\pr \vPhi^{(2)}}{\pr t} 
= \left({\cal L}^{(0)} \vPhi^{(2)}\right)^{[0]}
+   \left[ \vu_c e^{iq_c x + i \omega_c t} 
\left( v\frac{\pr A}{\pr \xi_1} -\frac{\pr A}{\pr \tau_1} \right)
+ \vu_c^* e^{-iq_c x- i \omega_c t} 
\left( v \frac{\pr A^*}{\pr \xi_1} -\frac{\pr A^*}{\pr \tau_1} \right)
\right] 
+ \left( {\cal N}_2[\vPhi]\right)^{[2]} \quad . 
\ee
The nonlinear term is easily evaluated with the help of
eqs.(\ref{bha}) and (\ref{bi})
\be{cd}
\left({\cal N}_2[\vPhi]\right)^{[2]} = -
2 \mL(0) \Ga |A|^2
- [\mL(2 q_c) - 2 i \omega_c \id]
\Gb e^{2i q_c x + 2 i \omega_c t}  A^2 
-
[\mL(- 2 q_c) + 2 i \omega_c \id]
\Gb^* e^{-2 i q_c x- 2 i \omega_c t} (A^*)^2 \quad ,
\ee
if the abbreviations (\ref{bmc}) and (\ref{bmd}) are used.
We have to check the
secular condition (\ref{fd}),
to determine the solution $\vPhi^{(2)}$. 
Only the modes 
$\pm e^{i q_c x + i \omega_c t}$ contribute, 
because either the wavenumber or the frequency do not vanish. 
The evaluation of this condition yields
\be{ch}
0 = \left( \frac{\pr }{\pr \tau_1} -v \frac{\pr }{\pr \xi_1}\right) A 
\quad .
\ee
As a consequence the second and third term on the right hand side
of eq.(\ref{cg}) vanish. For the solution $\vPhi^{(2)}$ one obtains
discarding transients (cf.~eq.(\ref{fb}))
\be{ci}
\vPhi^{(2)} = 2 \Ga |A|^2 +
\Gb e^{2i q_c x + 2 i \omega_c t} A^2 + 
\Gb^* e^{-2 i q_c x- 2 i \omega_c t} (A^*)^2 
+ \vu_c e^{i q_c x + i \omega_c t} B(\tau_1,\ldots,\xi_1,\ldots)+
\vu_c^* e^{-i q_c x -i \omega_c t} B^*(\ldots) 
\quad ,
\ee
where the constant of integration may of course depend on the
slower scales.

To third order in $\eps$ the equation of motion (\ref{ba}) reads
\bea{cj}
& & \frac{\pr \vPhi^{(3)}}{\pr t} 
+\left( \frac{\pr \vPhi^{(2)}}{\pr t}\right)^{[1]} 
+ \vu_c e^{i q_c x + i\omega_c t} 
\frac{\pr A}{\pr \tau_2} + \vu_c^* e^{-i q_c x- i \omega_c t} 
\frac{\pr A^*}{\pr \tau_2} \nonumber\\
&=& {\cal L}^{(0)} \vPhi^{(3)}+
\left({\cal L}^{(0)} \vPhi^{(2)}\right)^{[1]}+
\left({\cal L}^{(0)} \vu_c e^{iq_c x + i\omega_c t} A\right)^{[2]} +
\left({\cal L}^{(0)} \vu_c^* e^{-iq_c x- i \omega_c t} 
A^*\right)^{[2]} \\
&+ & \left({\cal L}^{(2)} \vu_c e^{iq_c x+ i \omega_c t }\right) \cdot A +
\left({\cal L}^{(2)} \vu_c^* e^{-iq_c x - i \omega_c t}\right) \cdot A^*
+ \left( {\cal N}_2[\vPhi]\right)^{[3]}+
\left( {\cal N}_3[\vPhi]\right)^{[3]} \quad . \nonumber
\eea
From eq.(\ref{ci}) we have
\be{ck}
\left( \frac{\pr \vPhi^{(2)}}{\pr t}\right)^{[1]}
= \vu_c e^{i q_c x + i \omega_c t} \frac{\pr B}{\pr \tau_1}
+ \vu_c^* e^{-i q_c x -i \omega_c t} \frac{\pr B^*}{\pr \tau_1} + \cdots
\quad .
\ee
Here and in the remaining part of this section
$\cdots$ indicate Fourier--modes (with wave vector $\pm 2 q_c, \pm 3 q_c$ or
frequency $\pm 2 \omega_c, \pm 3 \omega_c$), which will not contribute 
to the secular condition.
In the same way we obtain using eq.(\ref{cf})
\be{cl}
\left({\cal L}^{(0)} \vPhi^{(2)} \right)^{[1]}
= - i \mL'(q_c) \vu_c e^{i q_c x +i \omega_c t}
\frac{\pr B}{\pr \xi_1} -
i \mL'(-q_c) \vu_c^* e^{-i q_c x -i \omega_c t}
\frac{\pr B^*}{\pr \xi_1}
+\cdots
\ee
With the identity analogous to eq.(\ref{ce})
\be{cm}
\left( \frac{\pr^{\alpha}}{\pr x^{\alpha}} e^{i q_c x+ i \omega_c t} 
A\right)^{[2]} 
= \alpha (i q_c)^{\alpha-1 } 
e^{i q_c x+ i \omega_c t} \frac{\pr A}{\pr \xi_2} 
+
\frac{\alpha(\alpha-1)}{2}
(i q_c)^{\alpha-2} e^{i q_c x+ i \omega_c t} \frac{\pr^2 A}{\pr \xi_1^2}
\ee
one has
\be{cn}
\left({\cal L}^{(0)} \vu_c e^{i q_c x + i \omega_c t} A\right)^{[2]} =
-i \mL'(q_c) \vu_c e^{i q_c x+ i \omega_c t} \frac{\pr A}{\pr \xi_2}
+ \frac{(-i)^2}{2}  \mL''(q_c) \vu_c 
e^{i q_c x + i\omega_c t}
\frac{\pr^2 A}{\pr \xi_1^2} \quad .
\ee
For the evaluation of the nonlinear contributions we again keep in mind
that only the resonant Fourier--modes have to be considered. 
Then
\bea{co}
\left({\cal N}_2[\ul{\Phi}]\right)^{[3]}
&=& 2
\sum_{\alpha\beta} 
C^{(\alpha\beta)} \left\{ (i q_c)^{\alpha} \vu_c 
e^{i q_c x + i \omega_c t} A +
(-i q_c)^{\alpha} \vu_c^* e^{-i q_c x- i \omega_c t} 
A^*,
\left( \frac{\pr^{\beta} \vPhi^{(2)}}{\pr x^{\beta}}\right)^{[0]} 
\right\}+ \cdots \nonumber\\
&=& \tG e^{i q_c x+ i \omega_c t} |A|^2 A + 
\tG^* e^{-i q_c x- i \omega_c t} |A|^2 A^* 
+ \cdots \quad ,
\eea
where for the evaluation eq.(\ref{ci}) and the abbreviation
(\ref{bmb})
have been used. In the same way one obtains
\be{cp}
\left( {\cal N}_3[\vPhi]\right)^{[3]} =
\ul{\Delta} e^{i q_c x + i \omega_c t} |A|^2 A +
\ul{\Delta}^* e^{-i q_c x- i \omega_c t} |A|^2 A^* +\cdots
\ee
using the abbreviation (\ref{bme}).

If we now collect eqs.(\ref{ck}), (\ref{cl}), and (\ref{cn})
the evolution equation (\ref{cj}) reads
\bea{cq}
\frac{\pr \vPhi^{(3)}}{\pr t} &=& 
\left({\cal L}^{(0)} \vPhi^{(3)}\right)^{[0]}
+\vu_c e^{i q_c x + i \omega_c t} \left(
v \frac{\pr B}{\pr \xi_1} -\frac{\pr B}{\pr \tau_1}\right)
+
\vu_c^* e^{-i q_c x -i \omega_c t} \left(
v \frac{\pr B^*}{\pr \xi_1} - \frac{\pr B^*}{\pr \tau_1} \right)
\nonumber \\
&+&
\vu_c e^{i q_c x + i \omega_c t} \left(
v \frac{\pr A}{\pr \xi_2} -\frac{\pr A}{\pr \tau_2}\right)
+
\vu_c^* e^{-i q_c x -i \omega_c t} \left(
v \frac{\pr A^*}{\pr \xi_2} - \frac{\pr A^*}{\pr \tau_2} \right)
\nonumber\\
&-& \frac{1}{2} \left.\frac{d^2 \lambda^{(\nu_c)}(k)}{dk^2}\right|_{k=q_c} 
\vu_c e^{i q_c x + i\omega_c t}
\frac{\pr^2 A}{\pr \xi_1^2}
-
\frac{1}{2} \left.\frac{d^2 \lambda^{(\nu_c) *}(k)}{dk^2}\right|_{k=q_c} 
\vu_c^* e^{- i q_c x - i\omega_c t}
\frac{\pr^2 A^*}{\pr \xi_1^2}
\nonumber\\
&+ & \left({\cal L}^{(2)} \vu_c e^{iq_c x+ i \omega_c t }\right) \cdot A +
\left({\cal L}^{(2)} \vu_c^* e^{-iq_c x - i \omega_c t}\right) \cdot A^*
\nonumber\\
&+& \left( {\cal N}_2[\vPhi]\right)^{[3]}+
\left( {\cal N}_3[\vPhi]\right)^{[3]} + \cdots \quad ,
\eea
where the nonlinear terms are given by eqs.(\ref{co}) and (\ref{cp}). 
In addition we have used eq.(\ref{cfa}) 
and the analogous relation for the 
second derivative. 

Only the terms
written explicitly will contribute to the secular condition, 
for the reasons mentioned above. Eq.(\ref{fd}) yields
\be{cr}
0=
\left( v \frac{\pr B}{\pr \xi_1} -\frac{\pr B}{\pr \tau_1}\right)
+
\left( v \frac{\pr A}{\pr \xi_2} -\frac{\pr A}{\pr \tau_2}\right)
+ D  \frac{\pr^2 A}{\pr \xi_1^2}
+ \eta A + r |A|^2 A
\ee
if one takes the abbreviations (\ref{blb}), (\ref{bma}), and (\ref{bn})
into account and recalls, that ${\cal L}^{(2)}$ is invariant with respect to
translations in space and time.

Finally one has to separate the amplitudes $A$ and $B$ from each other. For 
that purpose  apply $(\pr/\pr \tau_1 - v \pr/\pr \xi_1)$ to eq.(\ref{cr}).
Then
\be{cs}
0= \left( \frac{\pr }{\pr \tau_1} - v \frac{\pr }{\pr \xi_1} 
\right)^2 B
\ee
holds,
if we use secular condition (\ref{ch}) of the preceding order.
Since the general solution of this equation is given by
$B=f_0(\xi_1+v\tau_1) + (\xi_1-v \tau_1) f_1(\xi_1+ v \tau_1)$, but
$B$ must not contain a secular contribution, one has $f_1\equiv 0$.
Hence
\be{ct}
0= \left( \frac{\pr }{\pr \tau_1} - v \frac{\pr }{\pr \xi_1} 
\right) B \quad ,
\ee
and eq.(\ref{cr}) results in eq.(\ref{bk}).
\section{Discussion}\label{sec4}
The analysis of the preceding section has shown, that 
the dynamics beyond an instability of a single mode
is generically described by a complex Ginzburg--Landau equation, 
if the underlying dynamics is autonomous. Since the solutions of this 
reduced equation
are bounded for $\Re D >0$ and $\Re\, r <0$ (cf.~\cite{Temam,Collet}), 
the motion
is correctly described at least on time scales $t\sim \eps^{-2}$.
It is not the aim of this article to
go into the details of the properties of the amplitude equation,
which are itself a field of current research. Let us only mention that the
change in sign of $\Re D$ or $\Re\, r$ are higher--codimension
instabilities, i.e.~a kind of Eckhaus instability and the transition from
super-- to sub--critical behaviour respectively.

Taking symmetries of the underlying equations of motion into account,
the apparently complicated expressions for the coefficients of the
amplitude equation may be simplified considerably.
A frequent constraint in this direction is a system, being symmetric with
respect to space--inversion. Let us restrict therefore the 
following discussion to this case. As
the main consequence of the symmetry the matrix (\ref{bdd}) determining
the stability is always real. Hence one real or two complex conjugated 
critical eigenvalues for the unique critical mode occurs.

Consider first the case of a non--vanishing
critical wave number $q_c\neq 0$, which is sometimes called a
soft--mode instability. Then the restriction to one
critical mode implies  a vanishing frequency, 
even for wave-numbers in a neighbourhood of $q_c$.
Hence the derivatives of the spectrum and the eigenvectors are real
quantities. But then the convective velocity vanishes and the
remaining coefficients of the amplitude equation 
(cf.~eqs.(\ref{blb})--(\ref{bme})) are real, since only derivatives 
of even order occur.

If the critical wave vector vanishes, the frequency must be finite
$\omega_c \neq 0$. Such a situation is sometimes called a
hard--mode instability. Since the eigenvectors are now complex, 
the coefficients do not reduce to real numbers even in the presence of
inversion symmetry. But the convective velocity (\ref{bla}) vanishes
for that reason. In addition, since $q_c=0$, only the spatially homogeneous
contributions, i.e.~one summand, enter the formulas for the cubic
coefficient (cf.~eqs.(\ref{bma})--(\ref{bme})). 
Of course $r$ reduces to the expression known from the
simple Hopf bifurcation (cf.~\cite[p.152f]{GuHo}).
%
\section{Turing--Hopf Instability}\label{sec7}
As stated in the previous 
section soft-- and hard--mode instabilities are
typical, i.e.~of codimension one, in systems which are symmetric with respect
to space--inversion. The corresponding codimension two situation, 
where both instabilities occur simultaneously has recently attracted 
considerable interest even from the experimental point of view \cite{DDB}. 
For that reason we think it is useful to present the general analytical 
expressions for the coefficients of the corresponding coupled amplitude 
equations, which govern the dynamics beyond the instability.

Again we start our analysis from the general autonomous evolution equation 
(\ref{ba}) but impose in addition the inversion symmetry. 
Since a soft-- and a hard--mode become unstable simultaneously
we presuppose that instead of eq.(\ref{be}) the spectrum obeys
$\lambda^{(\nu_c)}(q_c)=0$ and $\lambda^{(\nu_c)}(0)=i \omega_c$. 
For the weakly nonlinear analysis one has to take both modes into account,
where the amplitudes may depend on the slower spatial and temporal scales
\be{ia}
\vPhi(x,t) = \eps \left[ \vu^S_c e^{i q_c x } 
A^S + \vu_c^S e^{-i q_c x } A^{S\,*} +  \vu_c^H e^{ i \omega_c t} 
A^H  + 
\vu_c^{H\,*} e^{ - i\omega_c t} A^{H\,*} \right] 
+ \eps^2 \vPhi^{(2)} + \eps^3 \vPhi^{(3)}+ \odr(\eps^4) \quad .
\ee
Since $\mL(q_c)$ is real the vector part $\vu_c^S:=\vu_{q_c}^{(\nu_c)}$ 
of the soft mode is real too.
In analogy to section \ref{sec2} the first order leads to the instability
condition. For the second order we obtain
\be{ib}
\frac{\pr \vPhi^{(2)}}{\pr t} 
= \left({\cal L}^{(0)} \vPhi^{(2)}\right)^{[0]} 
 -   \vu^S_c e^{iq_c x  } \frac{\pr A^S}{\pr \tau_1}  
- \vu_c^S e^{-iq_c x } 
 \frac{\pr A^{S\,*}}{\pr \tau_1}  
-    \vu^H_c e^{ i \omega_c t} \frac{\pr A^H}{\pr \tau_1}  
- \vu_c^{H\,*} e^{ - i \omega_c t}  \frac{\pr A^{H\,*}}{\pr \tau_1}  
+ \left( {\cal N}_2[\vPhi]\right)^{[2]}  
\ee
if we again make use of the fact, that the eigenvectors can be assumed to be
independent of the wavenumber, and that the first derivatives of the 
eigenvalues vanish by inversion symmetry. That part of eq.(\ref{ib}) which is 
linear in the amplitudes is nothing else but the identical copy of 
eq.(\ref{cg}) for the two modes. The nonlinear part reads
\bea{ic}
\left({\cal N}_2[\vPhi]\right)^{[2]} 
&=& -
2 \mL(0) \Ga^S |A^S|^2
- \mL(2 q_c) \Gb^S e^{2i q_c x  }  (A^S)^2 
- \mL(- 2 q_c)  \Gb^{S\,*} e^{-2 i q_c x } (A^{S\,*})^2 \nonumber\\
&-&
2 \mL(0) \Ga^H |A^H|^2
- [\mL(0) - 2 i \omega_c \id]
\Gb^H e^{  2 i \omega_c t}  (A^H)^2  - 
[\mL(0) + 2 i \omega_c \id]
\Gb^{H\,*} e^{ - 2 i \omega_c t} (A^{H\,*})^2 \nonumber\\
& -& 2 [\mL(q_c)-i \omega_c \id] \Gc e^{i q_c x + i \omega_c t} A^H A^S
- 2 [\mL(q_c)+i \omega_c \id] \Gc^* e^{-i q_c x - i \omega_c t} A^{H\,*} 
A^{S\,*} \nonumber\\
&-& 2 [\mL(q_c)-i \omega_c \id] \Gc e^{-i q_c x + i \omega_c t} A^H A^{S\,*}
- 2 [\mL(q_c)+i \omega_c \id] \Gc^* e^{i q_c x - i \omega_c t} A^{H\,*} 
A^S \quad .
\eea
The first two lines coincide with the corresponding expression from the
codimension one case (cf.~eq.(\ref{cd})). 
The coefficients $\ul{\Gamma}_{a/b}^S$
and  $\ul{\Gamma}_{a/b}^H$ are given by eqs.(\ref{bmc}) and (\ref{bmd})
evaluated for the soft-- and hard--mode respectively. The remaining terms
describe the interaction of the modes with the coefficient
being given by
\be{id}
\Gc := - \frac{1}{
\mL( q_c) -  i \omega_c \id}
\sum_{\beta} (i q_c)^{\beta} 
C^{(0 \beta)}\left\{\vu_c^H,\vu_c^S\right\} \quad.
\ee
Since two modes are critical, the secular condition (\ref{fd}) has to be
evaluated twice and yields
\be{ie}
0 = \frac{\pr A^S }{\pr \tau_1}, \qquad
0 = \frac{\pr A^H }{\pr \tau_1} \quad.
\ee
The the solution of second order, discarding transients, reads
\bea{if}
\vPhi^{(2)} &=&
 2 \Ga^S |A^S|^2 +
\Gb^S e^{2i q_c x  } (A^S)^2 + 
\Gb^{S\,*} e^{-2 i q_c x } (A^{S\,*})^2 
+ \vu_c^S e^{i q_c x  } B^S +
\vu_c^{S\,*} e^{-i q_c x  } B^{S\,*} \nonumber\\
&+&
2 \Ga^H |A^H|^2 +
\Gb^H e^{  2 i \omega_c t} (A^H)^2 + 
\Gb^{H\,*} e^{ - 2 i \omega_c t} (A^{H\,*})^2  
+ \vu^H_c e^{ i \omega_c t} B^H +
\vu_c^{H\,*} e^{ -i \omega_c t} B^{H\,*}  \nonumber\\
&+ & 2 \Gc e^{i q_c x + i \omega_c t} A^H A^S
+ 2 \Gc^* e^{-i q_c x - i \omega_c t} A^{H\,*} A^{S\,*}
+  2 \Gc e^{-i q_c x + i \omega_c t} A^H A^{S\,*}
+ 2 \Gc^* e^{i q_c x - i \omega_c t} A^{H\,*} A^S   \quad.
\eea
where the constants of integration $B^S$ and $B^H$ depend on the slower
scales. At third order we now obtain 
\bea{ig}
\frac{\pr \vPhi^{(3)}}{\pr t} &=& 
\left({\cal L}^{(0)} \vPhi^{(3)}\right)^{[0]}
-\vu^S_c e^{i q_c x }  \frac{\pr B^S}{\pr \tau_1} 
- \vu_c^S e^{-i q_c x  }  \frac{\pr B^{S\,*}}{\pr \tau_1}  
- \vu^S_c e^{i q_c x  } \frac{\pr A^S}{\pr \tau_2}
- \vu_c^S e^{-i q_c x  } \frac{\pr A^{S\,*}}{\pr \tau_2}  
\nonumber\\
&-& \frac{1}{2} \left.\frac{d^2 \lambda^{(\nu_c)}(k)}{dk^2}\right|_{k=q_c} 
\vu_c^S e^{i q_c x  }
\frac{\pr^2 A^S}{\pr \xi_1^2}
-
\frac{1}{2} \left.\frac{d^2 \lambda^{(\nu_c)}(k)}{dk^2}\right|_{k=q_c} 
\vu_c^S e^{- i q_c x  } \frac{\pr^2 A^{S\,*}}{\pr \xi_1^2}
\nonumber\\
&+ & \left({\cal L}^{(2)} \vu^S_c e^{iq_c x  }\right) \cdot A^S +
\left({\cal L}^{(2)} \vu_c^S e^{-iq_c x  }\right) \cdot A^{S\,*}
\nonumber\\
&-&\vu^H_c e^{ i \omega_c t}  \frac{\pr B^H}{\pr \tau_1} -
\vu_c^{H\,*} e^{ -i \omega_c t} \frac{\pr B^{H\,*}}{\pr \tau_1}  
-
\vu^H_c e^{ i \omega_c t}  \frac{\pr A^H}{\pr \tau_2} -
\vu_c^{H\,*} e^{ -i \omega_c t} \frac{\pr A^{H\,*}}{\pr \tau_2}  
\nonumber\\
&-& \frac{1}{2} \left.\frac{d^2 \lambda^{(\nu_c)}(k)}{dk^2}\right|_{k=0} 
\vu^H_c e^{ i\omega_c t}
\frac{\pr^2 A^H}{\pr \xi_1^2}
-
\frac{1}{2} \left.\frac{d^2 \lambda^{(\nu_c) *}(k)}{dk^2}\right|_{k=0} 
\vu_c^{H\,*} e^{ - i\omega_c t} \frac{\pr^2 A^{H\,*}}{\pr \xi_1^2}
\nonumber\\
&+ & \left({\cal L}^{(2)} \vu^H_c e^{ i \omega_c t }\right) \cdot A^H +
\left({\cal L}^{(2)} \vu_c^{H\,*} e^{ - i \omega_c t}\right) \cdot A^{H\,*}
\nonumber\\
&+& \left( {\cal N}_2[\vPhi]\right)^{[3]}+
\left( {\cal N}_3[\vPhi]\right)^{[3]} + \cdots \quad .
\eea
This apparently complicated expression contains the 
linear part of the two modes which is already known from the codimension one
analysis (cf.~eq.(\ref{cq})). The nonlinear contributions read 
\bea{ih}
\left({\cal N}_2[\ul{\Phi}]\right)^{[3]}
&=& \tG^S e^{i q_c x } |A^S|^2 A^S + 
\tG^{S\,*} e^{-i q_c x } |A^S|^2  A^{S\,*} 
+ \tG^H e^{ i \omega_c t} |A^H|^2 A^H + 
\tG^{H\,*} e^{ - i \omega_c t} |A^H|^2 A^{H\,*} \nonumber\\
&+& \ul{\gamma}^S e^{i q_c x} |A^H|^2 A^S
+ \ul{\gamma}^{S\,*} e^{-i q_c x} |A^H|^2 A^{S\,*}  
+ \ul{\gamma}^H e^{i \omega t} |A^S|^2 A^H
+ \ul{\gamma}^{H\,*} e^{-i \omega t} |A^S|^2 A^{H\,*} + \cdots
\eea
if we restrict the presentation to terms which will contribute to the
secular condition. Again the first line describes the contribution 
already known from the
analysis presented above,  
with the coefficients $\tG^S$ and $\tG^H$ being determined
by eqs.(\ref{bmb})--(\ref{bmd}) evaluated for the soft-- and 
hard--mode. The new coefficients which mediate the interaction are given by 
\bea{ii}
\ul{\gamma}^S &:=& 2 \sum_{\beta} (i q_c)^{\beta}
C^{(0 \beta)}\{ \vu_c^{H\,*}, 2 \Gc \} +
2 \sum_{\beta} (i q_c)^{\beta}
C^{(0 \beta)}\{ \vu_c^H, 2 \Gc^* \}
+ 2 \sum_{\alpha} (i q_c)^{\alpha} C^{(\alpha 0)}\{\vu_c^S,
2 \Ga^H\} \qquad (\in {\BbbR}^N) \label{iia} \\
\ul{\gamma}^H &:=& 2 \sum_{\alpha \beta} (-i q_c)^{\alpha}
(i q_c)^{\beta} C^{(\alpha \beta)}\{ \vu_c^S, 2 \Gc \} +
2 \sum_{\alpha \beta} (i q_c)^{\alpha}
(-i q_c)^{\beta} C^{(\alpha \beta)}\{ \vu_c^S, 2 \Gc \}
+ 2 C^{(0 0)}\{\vu_c^H, 2 \Ga^S\}
\label{iib} \quad.
\eea
In the same way the other term reads
\bea{ij}
\left( {\cal N}_3[\vPhi]\right)^{[3]}
&=&
\ul{\Delta}^S e^{i q_c x  } |A^S|^2 A^S +
\ul{\Delta}^{S\,*} e^{-i q_c x } |A^S|^2 A^{S\,*}   
+ \ul{\Delta}^H e^{ i \omega_c t} |A^H|^2 A^H +
\ul{\Delta}^{H\,*} e^{ - i \omega_c t} |A^H|^2 A^{H\,*}\nonumber\\ 
&+& \ul{\delta}^S  e^{i q_c x} |A^H|^2 A^S +
\ul{\delta}^{S\,*}  e^{-i q_c x} |A^H|^2 A^{S\,*} 
+ \ul{\delta}^H  e^{i \omega_c t} |A^S|^2 A^H +
\ul{\delta}^{H\,*}  e^{-i \omega_c t} |A^S|^2 A^{H\,*}
+\cdots \quad ,
\eea
where the interaction coefficients are
\bea{ik}
\ul{\delta}^S &:=&
6 \sum_{\gamma} 
(i q_c)^{\gamma} D^{(0 0 \gamma)} \{ \vu_c^H, \vu_c^{H\,*}, \vu_c^S \}
\qquad (\in {\BbbR}^N) \label{ika} \\
\ul{\delta}^H &:=&
6 \sum_{\beta \gamma}
(-i q_c)^{\beta} (i q_c)^{\gamma}
D^{(0 \beta \gamma)} \{ \vu_c^H, \vu_c^S, \vu_c^S \}
\label{ikb} \quad .
\eea
If one evaluates the secular conditions arising from 
eqs.(\ref{ig}), (\ref{ih}), and (\ref{ij}) and separates the amplitudes
$A$ and $B$ one finally obtains the coupled set of amplitude equations
\bea{il}
\frac{\pr A^S}{\pr \tau_2}
&=& \eta^S A^S + r^S |A^S|^2 A^S +
s^S |A^H|^2 A^S +
D^S \frac{\pr A^S}{\pr \xi_1^2}  \label{ila} \\
\frac{\pr A^H}{\pr \tau_2} 
&=& \eta^H A^H + r^H |A^H|^2 A^H +
s^H |A^S|^2 A^H +
D^H \frac{\pr A^H}{\pr \xi_1^2}  \quad , \label{ilb}
\eea
with the coupling coefficients being given by
\be{im}
s^S = \frac{ ( \ul{v}_c^S | \ul{\gamma}^S+\ul{\delta}^S) }
{ (\ul{v}_c^S | \ul{u}_c^S ) } \quad (\in {\BbbR}), \qquad
s^H = \frac{ ( \ul{v}_c^H | \ul{\gamma}^H+\ul{\delta}^H) }
{ (\ul{v}_c^H | \ul{u}_c^H ) } 
\ee
and eqs.(\ref{iia}), (\ref{iib}), (\ref{ika}), (\ref{ikb}). 
The remaining coefficients are of course
not changed compared to the codimension one case and are
given by eqs.(\ref{blb})--(\ref{bn}) evaluated for the soft--
and the hard--mode. 

We have obtained the well known set of coupled real and
complex Ginzburg--Landau equations \cite{Kida}, but with the general 
and closed 
analytical formulas for its coefficients. The simplicity of the whole 
derivation for this quite complicated instability again justifies the 
slight amount of formalism presented in the preceding sections.
\section{Explicitly Time--Dependent Equations}\label{sec5}
Let us now return to the case of a single unstable mode and
consider the case that the equations of motion (\ref{ba}),
that means the linear operator ${\cal L}$ and the nonlinear contributions
${\cal N}$, possess an additional explicit 
time--dependence with period $T$ and frequency $\Omega=2\pi/T$
respectively. Such a dependence may
arise in the presence of external time--dependent fields. But one should also
keep in mind, that a time--dependency may be introduced also 
by the trivial state,
if the equation of motion is cast into the form (\ref{ba}).
For the derivation of an 
amplitude equation the steps of the
preceding section can be applied. The calculation is quite similar
and we use the same symbols, even if they are now $T$--periodic functions
of time. However, two main formal differences occur. On the one hand
the stability of the trivial state is governed by a Floquet--
instead of an eigenvalue problem, that means that the eigenvectors itself are
$T$--periodic in time and obey
\bea{da}
\mu^{(\nu)}(k) \vu_k^{(\nu)}(t) +
\dot{\vu}_k^{(\nu)}(t) &=&
\mL (k,t) \vu_k^{(\nu)}(t) \label{daa} \\
\vv^{(\nu)*}_k(t)  \mu^{(\nu)}(k) 
- \dot{\vv}_k^{(\nu)*}(t) &=&
\vv_k^{(\nu)*}(t)  \mL(k,t) \label{dab}\quad .
\eea
Here $\mL(k,t)$ is defined by eq.(\ref{bdd}) with $\mL_{\alpha}$ being 
explicitly time--dependent, and the imaginary part of
the Floquet--exponents is restricted to the first Brillouin zone, 
$\Im \mu^{(\nu)}(k) \in [-\Omega/2, \Omega/2]$.
On the other hand the matrix coefficients in the abbreviations (\ref{bmc}) and
(\ref{bmd}), which are actually some propagators of the linear equation, 
are replaced by the corresponding quantity of the time--dependent equation.
We will present the details below.

The main physical difference comes from the fact, that one has to pay 
attention to strong resonances, which are known to be crucial already
in the corresponding spatially homogeneous situation \cite{Arnold}. 
For that reason it is
worthwhile to consider the explicitly time--dependent case separately. 

We first consider the situation without strong resonances.
Let us assume that either the
critical wave number $q_c$ does not vanish or that the frequency
$\omega_c$ does not obey a strong resonance condition, i.e.
\be{db}
\omega_c \neq \pm \frac{\Omega}{2}, \pm \frac{\Omega}{3}, 
\pm \frac{\Omega}{4} \quad .
\ee
We now proceed along the lines of section \ref{sec3}. 
It is not necessary to write down all equations again, 
since almost all formal steps are identical. Therefore
only those expressions are made explicit which change. 
We insert eq.(\ref{bi}) into the equation of
motion, keeping in mind that the eigenvector is time--dependent, 
and compare order for order in $\eps$. 

To first order the Floquet--equation (\ref{daa}) is reproduced.
The equation of motion for the second order is identical to eq.(\ref{cg}),
where we again profit from the fact that 
$k$--independent eigenvectors can be assumed (cf.~appendix \ref{appB}).
Here $i v$ of course denotes the derivative of the critical Floquet--exponent
with respect to the wave number.
For use of latter reference we write down 
the nonlinear contribution explicitly
\bea{dc}
\left( {\cal N}_2[\vPhi]\right)^{[2]} &=&  2 
\sum_{\alpha \beta} (i q_c)^{\alpha} (-i q_c)^{\beta}
C^{(\alpha \beta)}_t\left\{\vu_c(t),\vu_c^*(t)\right\} |A|^2 \nonumber\\
&+&
\sum_{\alpha \beta} (i q_c)^{\alpha} (i q_c)^{\beta}
C^{(\alpha \beta)}_t\left\{\vu_c(t),\vu_c(t)\right\} 
 e^{2i q_c x + 2 i \omega_c t}  A^2 \nonumber\\
&+&
\sum_{\alpha \beta} (- i q_c)^{\alpha} (- i q_c)^{\beta}
C^{(\alpha \beta)}_t\left\{\vu^*_c(t),\vu^*_c(t)\right\} 
 e^{-2i q_c x - 2 i \omega_c t}  (A^*)^2 \quad ,
\eea
where the $T$--periodic time--dependencies are
indicated. Since either the wavenumber does not vanish or the
non--resonance condition (\ref{db}) holds, evaluation of the secular 
condition 
(\ref{fe}) leads again to eq.(\ref{ch}). Hence the solution discarding
transients is given by eq.(\ref{ci}) (cf. appendix \ref{appB})
but with $\Ga$ and $\Gb$ being replaced
by the $T$--periodic quantities 
\bea{dd}
\Ga(t) &:=& \int_{-\infty}^t \mU_{k=0}(t,t')
\sum_{\alpha \beta} (i q_c)^{\alpha} (-i q_c)^{\beta}
C^{(\alpha \beta)}_{t'}\left\{\ul{u}_c(t'),\ul{u}_c^*(t')\right\}\, dt' 
\label{dda} \\
&=& 
\left[\id -\mM_{k=0}\right]^{-1}
 \int_{t-T}^t \mU_{k=0}(t,t')
\sum_{\alpha \beta} (i q_c)^{\alpha} (-i q_c)^{\beta}
C^{(\alpha \beta)}_{t'}\left\{\ul{u}_c(t'),
\ul{u}_c^*(t')\right\}\, dt' \nonumber \\
\Gb(t) &:=& \int_{-\infty}^{t}
\mU_{2 q_c} (t,t') e^{-2 i \omega_c (t-t') }
\sum_{\alpha \beta} (i q_c)^{\alpha} (i q_c)^{\beta}
C^{(\alpha \beta)}_{t'}\left\{\ul{u}_c(t'),\ul{u}_c(t')\right\}\, dt'
\label{ddb}\\
&=&
\left[\id -\mM_{2 q_c} e^{-2 i \omega_c T} \right]^{-1}
\int_{t-T}^t \mU_{2 q_c}(t,t')
e^{-2 i \omega_c (t-t') }
\sum_{\alpha \beta} (i q_c)^{\alpha} (i q_c)^{\beta}
C^{(\alpha \beta)}_{t'}\left\{\ul{u}_c(t'),\ul{u}_c(t')\right\}\, dt'
\quad . \nonumber
\eea
The indefinite time--integrals have been reduced to definite ones
by using the Floquet--decomposition (\ref{ff}) and the periodicity of 
the remaining factors.

We proceed to the third order and obtain eq.(\ref{cq}) with the derivatives
of the eigenvalues being replaced by derivatives of the Floquet--exponents.
For convenience we repeat the explicit expression for the nonlinear 
contributions
\bea{dea}
\left( {\cal N}_2[\vPhi]\right)^{[3]} &=&
\tG(t) e^{i q_c x+ i \omega_c t} |A|^2 A +
\tG^*(t) e^{- i q_c x- i \omega_c t} |A|^2 A^* + \cdots \label{deb}\\
\left( {\cal N}_3[\vPhi]\right)^{[3]} &=&
\ul{\Delta}(t) e^{i q_c x+ i \omega_c t} |A|^2 A +
\ul{\Delta}^*(t) e^{-i q_c x- i \omega_c t} |A|^2 A^* + \cdots 
\label{dec}
\quad .
\eea
Here $\tG(t)$ and $\ul{\Delta}(t)$ are again given by eqs.(\ref{bmb}) and
(\ref{bme}) evaluated with the time dependent quantities. 
Transients and terms with wavenumber $\pm 2 q_c, \pm 3 q_c$ or
frequency $\pm 2 \omega_c, \pm 3 \omega_c$,
which do not contribute to the secular condition by virtue of the 
non--resonance
condition (\ref{db}), have been indicated by $\cdots$.
The evaluation of the secular condition (\ref{fe}) leads to 
eq.(\ref{cr}) and the subsequent considerations are the same as in 
section \ref{sec3}. Hence we
again obtain the amplitude equation (\ref{bk}). 
But since the integrand in the
secular condition is time--dependent, 
the integral survives and the coefficients
read
\bea{df}
v &=& \left. \Im \frac{d \mu^{(\nu_c)}(k)}{d k} \right|_{k=q_c}  
\label{dfa}\\
D &=& -\frac{1}{2} \left. \frac{ d^2 \mu^{(\nu_c)}(k)}
{d k^2}\right|_{k=q_c}\label{dfb}\\
r &=& \frac{ \int_0^T ( \ul{v}_c(t) | \tG(t)+\ul{\Delta}(t)) \, dt }
{ \int_0^T (\ul{v}_c(t) | \ul{u}_c(t) )\, dt }\label{dfc} \\
\eta &=& \frac{ q_c/(2\pi) \int_0^T \int_0^{2\pi/q_c} 
\left( \vv_c(t) e^{i q_c x}
| {\cal L}^{(2)}_t | \vu_c(t) e^{i q_c x}\right) \, dx\, dt} 
{ \int_0^T (\ul{v}_c(t) | \ul{u}_c(t) )\, dt} \label{dfd} \quad .
\eea
One might argue that the evaluation of these formulas is in general
impossible. But one should notice that one only needs the solution
of the time--dependent problem (\ref{fea}) for one period 
(cf.~eqs.(\ref{dda}) and (\ref{ddb})). This is at least numerically
a very simple task, so that the evaluation even for quite complicated
equations of motion can be performed on every computer. But also
analytical computations are possible, if the Floquet--problem can be
handled, e.g.~with a separate perturbation expansion.
\section{Strong Resonances}\label{sec6}
From the analysis of the preceding section it is obvious, that in cases
of strong resonances additional terms contribute to the secular condition.
We will discuss in the sequel the implications of
each case separately and therefore assume 
$q_c=0$ throughout this section.
\paragraph{Quartic Hopf--Hard--Mode Instability}
Consider the case, that the frequency of the marginally stable mode 
obeys\footnote{A pair of complex conjugate eigenvalues occurs 
$\omega^{(\nu_c)}(0)=\omega_c$ and $\omega^{(\nu_c')}(0)=-\omega_c$.
The branch $\nu_c'$ leads to the complex conjugate expressions.}
\be{ea}
\omega_c=\frac{\Omega}{4}\quad .
\ee
Since this condition puts an additional constraint on the instability,
this situation can be roughly classified as a codimension--two
instability. The analysis of the preceding section up to and including
the equation of third order (cf.~eq.(\ref{cq})) 
is valid. But now the terms with frequency $\pm 3 \omega_c$
also contribute to the secular condition. Hence for the evaluation of the
nonlinear contribution (cf.~eqs.(\ref{bha}), (\ref{bhc}),
(\ref{bi}), (\ref{ci}), (\ref{deb}),
(\ref{dec})) one has to considers these terms too
\bea{eb}
\left( {\cal N}_2[\vPhi]\right)^{[3]} &=&
\tG(t) e^{i \omega_c t} |A|^2 A +
\tG^*(t) e^{- i \omega_c t} |A|^2 A^* 
\nonumber\\
&+& \ul{\gamma}(t) 
\left(A^*\right)^3 e^{-3 i \omega t + i \Omega t}
+ \ul{\gamma}^* (t) A^3 e^{3 i \omega t - i \Omega t}
+ \cdots
\label{eba}\\
\left( {\cal N}_3[\vPhi]\right)^{[3]} &=&
\ul{\Delta}(t) e^{i \omega_c t} |A|^2 A +
\ul{\Delta}^*(t) e^{- i \omega_c t} |A|^2 A^* \nonumber\\
&+& \ul{\delta}(t) \left( A^*\right)^3 
e^{- 3 i \omega_c t + i \Omega t} +
\ul{\delta}^*(t) A^3 
e^{3 i \omega_c t - i \Omega t} + \cdots
\label{ebb}
\quad .
\eea
Here
\bea{ec}
\ul{\gamma}(t) &:=& 2 C^{(0 0)}_t\left\{\ul{u}_c^*(t),
\ul{\Gamma}_b^* (t)\right\} e^{- i \Omega t}\label{eca} \\
\ul{\delta}(t) &:=& D^{(0 0 0)}_t \left\{\ul{u}_c^*(t),
\ul{u}_c^*(t), \ul{u}_c^*(t) \right\} e^{- i \Omega t} \label{ecb}
\eea
denote the additional $T$--periodic coefficients\footnote{Since $q_c=0$
only the term $\alpha=\beta=\gamma=0$ contributes in eqs.(\ref{bmb}),
(\ref{bmc}), (\ref{bmd}), and (\ref{bme}).},
whereas $\cdots$ indicate those terms with frequency $\pm 2\omega_c$ which
do not contribute to the secular condition. Because of the resonance 
condition (\ref{ea}) the last summands in eqs.(\ref{eba}) and (\ref{ebb})
do not drop in 
eq.(\ref{fe}). Instead of eq.(\ref{bk}) one obtains the amplitude equation, 
after having separated as usual the amplitudes $A$ and $B$
\be{ed}
\left( \frac{\pr }{\pr \tau_2} - v \frac{\pr }{\pr \xi_2}\right) A 
= \eta A + r |A|^2 A + s \left(A^*\right)^3
+D \frac{\pr^2 A}{\pr \xi_1^2} \quad .
\ee
The additional coefficient is given by
\be{ee}
s = \frac{ \int_0^T \left( \vv_c(t) | \ul{\gamma}(t)+
\ul{\delta}(t)\right) \, dt }
{ \int_0^T \left( \vv_c(t) | \vu_c(t) \right) \, dt } \quad ,
\ee
whereas for the remaining quantities the formulas of the preceding 
section apply. In contrast to the usual hard--mode instability, 
the amplitude equation does not
possess a phase symmetry. Hence the complex phase of $\eta$ cannot be
eliminated, and one has a two--dimensional ''unfolding--parameter''.
\paragraph{Flip--Hard--Mode Instability}
Let the frequency obey
\be{ef}
\omega_c=\frac{\Omega}{2}\quad .
\ee
The corresponding Floquet--exponent is located at the boundary of the 
Brillouin
zone, that means the Floquet--multiplier is isolated and takes the
value $-1$. Such a value induces a period doubling bifurcation, 
which is of course a structurally stable bifurcation of codimension one,
so that the condition (\ref{ef}) does not imply an additional 
constraint for the bifurcation.
If one takes into account that $\mL(0,t)$ is a real matrix, then
the complex part can be eliminated in the eigenvalue equations
(\ref{daa}) and (\ref{dab}) using the abbreviations
\bea{eg}
\vu_c(t) =: e^{-i \omega_c t} \vhu_c (t),&\quad&
\vhu_c(t)=-\vhu_c(t+T)\in {\BbbR}^N\label{ega}\\
\vv_c^*(t)=: e^{i \omega_c t} \vhv_c(t),&\quad&
\vhv_c(t)=-\vhv_c(t+T)\in {\BbbR}^N\label{egb} \quad .
\eea
Here the real vector $\vhu_c(t)$ points into the direction of the 
centre manifold, which in this case is a M\"obius strip. As a consequence
eq.(\ref{bi}) simplifies to
\bea{eh}
\vPhi(x,t)&=&2 \eps \vhu_c (t) 
A_r(\tau_1,\tau_2,\ldots,\xi_1,\xi_2,\ldots) + 
\eps^2 \vPhi^{(2)} + \eps^3 \vPhi^{(3)}+ \cdots \nonumber \\ 
A_r &:=& \Re A \quad ,
\eea
so that the field is completely determined by real part of the amplitude.
This property is a direct consequence of the fact,
that the centre manifold in the homogeneous system is a one--dimensional real
manifold.  It is now quite simple to evaluate the consequences of 
eqs.(\ref{ef}), (\ref{ega}), and (\ref{egb}) for the amplitude equation.
To second order one has the result (\ref{cg}) and (\ref{dc}). 
But to the secular condition
(\ref{fe}) now both, the second and third summand on the right hand side
of eq.(\ref{cg}) contribute
\be{ei}
0 = \left(v
\frac{\pr }{\pr \xi_1} - \frac{\pr }{\pr \tau_1}\right)
\left( A+ A^*\right)
= 2 \left(v
\frac{\pr }{\pr \xi_1} - \frac{\pr }{\pr \tau_1}\right) A_r \quad .
\ee
In addition, if we insert eqs.(\ref{ega}) and (\ref{egb}) into
the definitions (\ref{dda}) and (\ref{ddb}) we have
\be{ej}
\Ga (t)= e^{2 i \omega_c t} \Gb (t)=
\int_{-\infty}^t \mU_{k=0}(t,t')
C^{(0 0)}_{t'}\left\{\vhu_c(t'),\vhu_c(t')\right\}\, dt' \in {\BbbR}^N
\quad ,
\ee
since the evolution matrix is a real quantity. Hence
the solution (\ref{ci}) of second order simplifies to
\be{ek}
\vPhi^{(2)} = 
\Ga (t) \left( A + A^* \right)^2 + \vhu_c(t) \left(B + B^*\right)
= 4 \Ga (t) A_r^2 + 2 \vhu_c(t) B_r \quad .
\ee
To third order we already have eq.(\ref{cq}), where all the 
cubic terms in the amplitude contribute, that means (\ref{eba}) and
(\ref{ebb}) apply. If we evaluate the coefficients (\ref{bmb}), (\ref{bme}),
(\ref{eca}), and (\ref{ecb}) using eqs.(\ref{ef}), (\ref{ega}), (\ref{egb}),
and (\ref{ej}) we obtain
\bea{el}
\left({\cal N}_2[\vPhi,t]\right)^{[3]}
&=& \hat{\tG}(t) \left(A + A^*\right)^3
+ \cdots \label{ela}\\
\left( {\cal N}_3[\vPhi,t]\right)^{[3]} &=& \hat{\ul{\Delta}}(t) 
\left(A + A^*\right)^3 + \cdots \label{elb}
\eea
where
\bea{em}
\hat{\tG}(t) &=& 
\frac{1}{3} e^{i \omega_c t} \tG(t) =
2 C^{(0 0)}_t \left\{ \vhu_c(t), \Ga(t) \right\} 
\in {\BbbR}^N \label{ema}\\
\hat{\ul{\Delta}}(t) &=& 
\frac{1}{3} e^{i \omega_c t} \ul{\Delta}(t)=
D^{(0 0 0)}_t \left\{ \vhu_c(t),\vhu_c(t),
\vhu_c(t)\right\} \in {\BbbR}^N \label{emb} \quad .
\eea
The secular condition (\ref{fe}) then leads to the
amplitude equation
\be{en}
\left( \frac{\pr }{\pr \tau_2} - v \frac{\pr }{\pr \xi_2}\right) A_r 
= \eta A_r + 4 r A_r^3
+D \frac{\pr^2 A_r}{\pr \xi_1^2} \quad ,
\ee
if the amplitude $B_r$ is separated as usual. Again the velocity and the
real diffusion coefficient are determined by the derivatives of the
spectrum (\ref{dfa}) and (\ref{dfb}), whereas the remaining quantities
are expressed in terms of the centre manifold coordinate
\bea{eo}
r &=&  \frac{ \int_0^T \left( \vhv_c(t) | \hat{\tG}(t)+
\hat{\ul{\Delta}}(t)
\right) \, dt }{ \int_0^T (\vhv_c(t) | \vhu_c(t) )  \, dt} \in {\BbbR}
\label{eoa}\\
\eta &=& \frac{ q_c/(2\pi) \int_0^T \int_0^{2\pi/q_c} 
\left( \vhv_c(t)
| {\cal L}^{(2)}_t | \vhu_c(t) \right) \, dx\, dt} 
{ \int_0^T (\vhv_c(t) | \vhu_c(t) )\, dt} \in {\BbbR} \label{eob} \quad .
\eea
Eq.(\ref{en}), which is entirely real including the amplitude, should not be
mixed up with the real Ginzburg--Landau equation. It is sometimes 
called a Fishers equation \cite{Fischer}.
\paragraph{Cubic Hopf--Hard--Mode Instability}
Yet there appeared only modifications in the amplitude equations,
but the general perturbation scheme was not influenced. 
This feature changes considerably for a third order degeneracy
in the Floquet--multipliers, i.e.
\be{ep} 
\omega_c=\frac{\Omega}{3}\quad .
\ee
It is evident from eqs.(\ref{cg}) and (\ref{dc}) that the nonlinearities
contribute to the secular condition.
Indeed we obtain\footnote{A linear and a
diffusive term can be introduced by using a different
scaling of the amplitude and the spatial coordinate with $\eps$. It does not
seem to change the subsequent considerations.}
\be{eq}
\left( \frac{\pr }{\pr \tau_1} -v \frac{\pr }{\pr \xi_1} \right) A
= \alpha \left( A^*\right)^2 \quad ,
\ee
with a coefficient $\alpha$ being determined by quadratic nonlinearities.
It is not difficult to show, that for almost all initial conditions 
eq.(\ref{eq}) yields an unbounded, that means a secular, solution. This 
feature is by no means amazing,
since a stabilising cubic term, known to be important 
in the normal form of the spatially homogeneous system, is missing.
One may cure this defect by tracing back to what is sometimes called a 
re-summation of the secular conditions \cite[p.318f]{Mann}. 
Although such an approach
seems to be quite common, it is difficult to estimate the validity of this
procedure. Especially the actual expansion parameter is unclear 
and one mixes the different scales in an uncontrolled manner. In fact, 
the formal concept of the multiple scale analysis implies, that
the secular conditions at each order have to
be satisfied separately. In this sense the codimension--two bifurcation
of this paragraph cannot be treated by the perturbation scheme.
\section{Conclusion}
It was shown by explicit calculation that for every
codimension--one instability
of a trivial state
the dynamics beyond the threshold is governed by a Ginzburg--Landau or
a Fishers equation. { Especially for the frequently met case of 
a soft--mode 
or a hard--mode instability in autonomous systems, there arises a unique 
expression for the coefficients of the amplitude equation, which covers
both the real as well as the complex Ginzburg--Landau equation.}
The reader may object that all these results can be
obtained simply from normal forms of ordinary differential equations
supplemented with symmetry considerations, and that its is
not necessary to go through the
explicit derivation. Alongside the disadvantage,
that such an approach is incapable to yield
numerical values for the coefficients, e.g.~to locate transitions from
super-- to sub--critical instabilities, one has to be careful concerning the
validity of such phenomenological amplitude equations.
It might happen, that in the stage of the derivation
secular conditions occur, which put severe constraints on the validity of 
the multiple scales approach. 
One such constraint, which in fact invalidates the
approach, was presented in section \ref{sec6} in conjunction with strong
resonances. A more prominent example is known in the context of
counter--propagating waves \cite{KnoLe}. Here the thorough approach leads
to a nonlocal reduced description, whereas the validity of
the phenomenological amplitude equations containing convection terms is
limited to a higher--codimension instability. A similar phenomenon is known in
the context of phase turbulence, described by the Kuramoto--Sivashinski
equation \cite{Kura76a}. Hence symmetry considerations and normal forms
of ordinary differential equations are very good guidelines for the resulting
amplitude equation, but they do not substitute a formal derivation.

The general and formal approach has shown that it is by no means essential to
implement certain properties of the perturbation expansion
with ad hoc assumptions. In fact all scales which seem to be 
superfluous drop by itself, which emphasises 
the consistency of the perturbation
expansion. One should however stress, that the whole physics 
(or mathematics, depending on the 
readers taste), is contained in the expansion (\ref{bi}) of the solution. 
All the remaining steps are just straightforward. Hence, 
by a change of this expansion it is obvious, that different situations, 
e.g.~higher--codimension instabilities, phase equations etc. can be also
handled for a general equation of motion. 
Since the whole formalism is quite simple,
it is indeed possible to treat higher--codimension bifurcations,
and to obtain the amplitude equation from the
basic equation of motion as demonstrated on the
example of the Turing--Hopf instability. Different cases, e.g.~the
degenerated soft--mode instability which require perturbation expansions
of higher order have been already treated
\cite{Matt}, and will be published elsewhere.

Finally the formal and general nature of the presented treatment 
may clearly { indicate the principal limits of the approach
by amplitude equations. Although it was not our purpose to touch
the asymptotic properties of the expansion 
from the mathematical point of view, 
our scheme probably contributes to this field as well, e.g.~one may 
construct a clear connection to the normal form theory of low 
dimensional dynamical
systems. But a thorough investigation will require methods which
are beyond the scope of this publication.}
\appendix
\section{Transformation Properties}\label{appA}
Suppose $\vu_k^{(\nu_c)}$ depends on $k$. We can perform a unitary 
transformation $\mR(k)$ depending continuously on $k$, so that
\be{ga}
\mR(k) \vu_k^{(\nu_c)}=\vu_c,\quad \mR(q_c)=\id
\ee
holds. The symmetry relation $\mR^*(k)=\mR(-k)$ may be 
imposed\footnote{We need the transformation as a formal tool only in
a neighbourhood of $|k|=|q_c|$. Hence we need not worry about
the global continuation.}.
We are now going to use this rotation to transform the full partial
differential equation (\ref{ba}) in order to obtain $k$--independent critical
eigenvectors. The main trick however is, that the final results
(cf. section \ref{sec2}) are scalars so that the transformation cancels in
these formulas. Hence the evaluation of these expressions based on 
the original equation is permitted and one may not take any
notice from the transformation.

To perform the transformation (\ref{ga}) on each Fourier--mode let us
define the real operator
\be{gb}
{\cal R}:= \mR\left( - i \frac{\pr}{\pr x} \right) \quad .
\ee
Its inverse can also been introduced since $\mR(k)$ is unitary.
With the transformed field
\be{gc}
\tvPhi(x,t) := {\cal R} \vPhi(x,t)
\ee
the equation of motion reads
\be{gd}
\frac{\pr \tvPhi}{\pr t} = {\cal R} {\cal L} 
{\cal R}^{-1} \tvPhi + {\cal R} {\cal N}[{\cal R}^{-1} \tvPhi]
=: \tL \tvPhi + \tN[\tvPhi] \quad .
\ee
We now apply the derivation of section \ref{sec3}, 
since by definition (\ref{ga})
the eigenvector $\tilde{\vu}_k^{(\nu_c)}$ of
\be{ge}
\ul{\ul{\tilde{L}}}(k):= \mR(k) \mL(k) \mR^{-1}(k)
\ee
does not depend on $k$. Hence we obtain the result of section \ref{sec2} 
but of course the coefficients (\ref{bla}), (\ref{blb}), (\ref{bma}),
and (\ref{bn}) are expressed in terms of the new quantities.
We have to show that the transformation
drops from this expression.

First the eigenvalues $\lambda^{(\nu)}(k)$ are independent of
the transformation ${\cal R}$, so that expression (\ref{bla}) and (\ref{blb})
can be evaluated in terms of the original quantities (\ref{ba}).
Second the transformed field (\ref{gc}) has an amplitude $\tilde{A}$
which may differ from the amplitude $A$ of the original field. But
if we insert the definition (\ref{bi}) into eq.(\ref{gc}), we clearly
observe, taking eq.(\ref{ga}) into account, that both amplitudes
coincide up to order $\eps$. 
Third the transformation drops from the matrix element (\ref{bn}), if
the definition for $\tL$ (cf.~eqs.(\ref{gc}) and (\ref{gd})) is inserted.
Finally we have to show, that the nonlinear coefficient can be evaluated from
the original quantities. 

For that purpose consider the transformed nonlinearity $\tN_2$, insert
a field consisting of two Fourier--modes $\tvpsi=\tvpsi_{q_1}\exp(i
q_1 x) + \tvpsi_{q_2} \exp(i q_2 x)+ c.c.$, and equate the Fourier--component
$\exp(i (q_1 + q_2) x)$. Then, according to the definition (\ref{bh})
and the transformation (\ref{gd}) one obtains
\be{gf}
\sum_{\alpha \beta} (i q_1)^{\alpha} (i q_2)^{\beta}
\tilde{C}^{(\alpha \beta)} \left\{ \tvpsi_{q_1}, \tvpsi_{q_2}\right\} =
\sum_{\alpha \beta} (i q_1)^{\alpha} (i q_2)^{\beta} \mR(q_1+q_2)
C^{(\alpha \beta)}\left\{ \mR^{-1}(q_1) \tvpsi_{q_1},
\mR^{-1}(q_2) \tvpsi_{q_2} \right\} \quad.
\ee
This formula tells us how the tensors change under the transformation.
We now repeatedly use this relation 
to evaluate the nonlinear coefficient
by choosing $q_1$, $q_2$,
$\tvpsi_{q_1}$, and $\tvpsi_{q_2}$ appropriately.
Consider the definition of $\tGa$ (cf.~eq.(\ref{bmc}))
and apply relation (\ref{gf}). Then
\be{gg}
\tGa = -\frac{1}{\tilde{\mL}(0)} \mR(0)
\sum_{\alpha \beta} (i q_c)^{\alpha} (-i q_c)^{\beta} 
C^{(\alpha \beta)} \left\{ \mR^{-1}(q_c) \tvu_c, \mR^{-1}(-q_c) \tvu_c^*
\right\} = \mR(0) \Ga
\ee
holds, where eqs.(\ref{ga}) and (\ref{ge}) were used
in the last step. In the same way
one obtains
\be{gh}
\tGb = \mR(2 q_c) \Gb \quad .
\ee
If we now insert both expression into the definition (\ref{bmb}) and
use again the transformation (\ref{gf}) appropriately we end up with
\bea{gi}
\ttG &=&
2 \sum_{\alpha} (i q_c)^{\alpha} \mR(q_c) C^{(\alpha 0)}\left\{
\mR^{-1}(q_c)
\tvu_c,
2 \mR^{-1}(0) \tGa \right\}
+ 2 \sum_{\alpha\beta} (2 i q_c)^{\beta} (- i q_c)^{\alpha}
\mR(q_c) C^{(\alpha \beta)}\left\{\mR^{-1}(-q_c) \tvu_c^*,
\mR^{-1}(2 q_c) \tGb \right\} \nonumber\\
&=& \tG \quad .
\eea
In the same way one obtains
\be{gj}
\tilde{\ul{\Delta}}= \ul{\Delta} \quad ,
\ee
so that the nonlinear coefficient $r$ can be evaluated from the
transformed as well as the original equation using 
the definitions (\ref{bma}), (\ref{bmb}), (\ref{bmc}), (\ref{bmd}),
and (\ref{bme}) in terms of the original quantities (\ref{ba})

The same approach can be used in the explicitly time--dependent case as well.
In that case one chooses a $T$--periodic transformation to make the
Floquet--vectors $k$--independent
\be{gk}
\mR(k,t) \vu_k^{(\nu_c)}(t)=\vu_c(t),\quad \mR(q_c,t)=\id \quad.
\ee
The only difference to the preceding considerations results from 
the time--dependence, which contributes an additional term to the
transformed linear operator (cf.~eq.(\ref{gd}))
\be{gl}
\tL_t =
{\cal R}_t {\cal L}_t 
{\cal R}^{-1}_t + \dot{\cal R}_t {\cal R}^{-1}_t \quad.
\ee
But it is exactly this property which ensures, that the propagators
(\ref{fea}),
occurring in the definitions (\ref{dda}) and (\ref{ddb}) 
of $\tGa(t)$ and $\tGb(t)$ respectively, 
transform according to
\be{gm}
\tilde{\mU}_k(t,t')= \mR(k,t) \mU_k(t,t') \mR(k,t') \quad .
\ee
Hence the transformation properties (\ref{gg}), (\ref{gh}),
(\ref{gi}), and (\ref{gj}) are valid in the
time--dependent case too.
\section{Secular Condition}\label{appB}
Let us determine the solution of the linear equation
\be{fa}
\frac{\pr \ul{\psi}}{\pr t} = {\cal L}^{(0)} \ul{\psi} +
\sum_k \ul{w}_k (t) e^{i k x} \quad ,
\ee
presupposing that
the linear operator has only stable modes except for one 
(cf. section \ref{sec2}) and that the inhomogeneous part, bounded in time,
is given by a finite sum of Fourier--modes. 

By considering each Fourier--component 
separately, the partial differential equation 
reduces to ordinary differential equations 
and the general solution is easily written down in terms of 
the matrix (\ref{bdd}) 
\be{fb}
\vpsi(x,t)= \vu_c e^{i q_c x + i \omega_c t} B + \sum_k \int_0^t
\exp\left[\mL(k) t' \right] \ul{w}_k(t-t') \, dt' e^{i k x} +
\cdots \quad .
\ee
Here $B$ denotes the constant of integration, and
$\cdots$ that part of the homogeneous solution, which 
corresponds to stable eigenvalues and leads to a transient only. 
Consider first the non--critical summands, that means $|k|\neq |q_c|$.
Then the integrals 
converge in the long time limit, since all eigenvalues have a
negative real part. For the marginally 
stable wavenumber $k=q_c$
one eigenvalue with a vanishing real part occurs, 
so that the integral may increase in time. 
By using e.g.~the spectral decomposition of the corresponding matrix
\be{fc}
\mL(q_c) = \sum_{\mu} \lambda^{(\mu)}(q_c) 
\frac{|\vu_{q_c}^{(\mu)})(\vv_{q_c}^{(\mu)}|}
{(\vv_{q_c}^{(\mu)}|\vu_{q_c}^{(\mu)})} \quad ,
\ee
it is evident from eq.(\ref{fb}) that a secular contribution increasing 
linearly in time is avoided if the relation
\be{fd}
0 = \lim_{\Theta\rightarrow\infty} \frac{1}{\Theta} \int_0^\Theta
e^{- i \omega_c t' } ( \vv_c | \ul{w}_{q_c} (t')) \, dt'
\ee
is fulfilled\footnote{Often such a relation is called a solvability condition
for the linear eq.(\ref{fa}). But such a term
requires the specification
of function spaces on which the operators are considered.}. 
In this case the solution in the stationary state is obtained by 
extending the upper limit of
the time integrals in eq.(\ref{fb}) to infinity. 

For a $T$--periodic explicitly 
time--dependent operator ${\cal L}^{0}_t$ 
similar considerations apply. The secular condition is e.g.~obtained
from eq.(\ref{fa}) by multiplication with the critical 
left--Floquet--eigenfunction $\vv_c(t) e^{i q_c x}$ (cf.~eq.(\ref{dab})).
One has to require
\be{fe}
0 = \lim_{\Theta\rightarrow\infty} \frac{1}{\Theta} \int_0^\Theta
e^{- i \omega_c t' } ( \vv_c(t') | \ul{w}_{q_c} (t')) \, dt'
\ee
in order to exclude linearly in time increasing contributions.
The solution of eq.(\ref{fa}) is again determined by the expression (\ref{fb})
with the matrix--exponential being replaced by the corresponding 
evolution operator
$\mU_k(t,t-t')$ of the time dependent system. The latter is determined by
\be{fea}
\frac{\partial \mU_k(t,t')}{\partial t} = \mL(k,t) \mU_k (t,t'),\quad
\mU_k(t',t')=\id \quad .
\ee
The solution in the stationary state is again obtained by extending the 
time integrals to infinity. This indefinite integrals
can be reduced to definite ones
by taking the Floquet--decomposition
\be{ff}
\mU_k(t+T,t')= \mM_k \mU_k(t,t')
\ee
into account. Here the constant 
matrix $\mM_k$ determines the Floquet--multipliers.
\section{Maxwell--Bloch equations}\label{appc}
{
For purely pedagogical purpose we present the evaluation
of the expressions given in section \ref{sec2} for the Laser instability,
which is governed by the Maxwell--Bloch equations \cite{Coullet}. 
Thus every formal step is performed explicitly. For the physical details
however the reader should consult the literature.

The basics equations of motion, which govern the evolution of the
complex valued envelopes $e$ and $p$ of the electric and the polarisation 
field as well as the deviation of the inversion $n$ from the pump level $r$,
read in dimensionless form
\bea{ha}
\frac{\partial e}{\partial t} &=& i a \frac{\partial^2 e}{\partial x^2}
-\sigma(e-p) \nonumber\\
\frac{\partial p}{\partial t} &=& -(1+i \Omega_0) p +(r-n)e\nonumber\\
\frac{\partial n}{\partial t} &=& -b n +(e^* p+e p^*)/2
\eea
where $x$ denotes the direction perpendicular to the beam.
Here the parameters $\sigma$, $\Omega_0$, and $b$ denote the cavity damping, 
the detuning, and the decay rate of the inversion in units of the 
dephasing rate. The parameter $a$ scales the diffraction term.
In the sequel we consider $\Omega_0<0$ and analyse the instability of
the trivial state $e=p=n=0$. 
Introducing the five real quantities
via
\be{hb}
e=e^{-i \Omega_0 t}(\Phi_1+i \Phi_2),\quad 
p=e^{-i \Omega_0 t}(\Phi_3+i \Phi_4),\quad n=\Phi_5
\ee
the equation of motion is cast into the form (\ref{ba})
\be{hc}
\frac{\partial \underline{\Phi}}{\partial t} =
\left(\begin{array}{ccccc}
- \sigma & - \Omega_0 - a \partial_x^2 & \sigma & 0 & 0 \\
\Omega_0 + a \partial_x^2 & - \sigma & 0 & \sigma & 0 \\
r & 0 & -1 & 0 & 0 \\
0 & r & 0 & -1& 0 \\
0 & 0 & 0 & 0 & -b \end{array} \right) \underline{\Phi}
+ \left( \begin{array}{c}
0 \\ 0 \\ -\Phi_1 \Phi_5 \\ -\Phi_2 \Phi_5 \\
\Phi_1 \Phi_3 + \Phi_2 \Phi_4 \end{array} \right)  \quad .
\ee
Since the nonlinearity is quadratic, the only nonvanishing contribution using
the notation (\ref{bg}), (\ref{bha}) and respecting the symmetry constraint
reads
\bea{hk}
C^{(00)}\{\underline{u},\underline{v}\} &=& \frac{1}{2}
\left(0, 0, -u_1 v_5-u_5 v_1, -u_2 v_5- u_5 v_2, \right. \nonumber\\
& & \left. \quad 
u_1 v_3 +u_3 v_1 + u_2 v_4 + u_4 v_2 \right)^T \quad .
\eea
The matrix (\ref{bdd}) is obtained from eq.(\ref{hc})
by replacing the differential $\partial_x$
by $i k$. Its characteristic polynomial reads
\be{hd}
0=\left[\lambda +b \right] \left[
\Omega_k^2 + \sigma^2 (1-r)^2 + 2 \lambda \{\Omega_k^2 + 
(1-r)(1+\sigma)\sigma\}
+ \lambda^2\{(1+\sigma)^2 + \Omega_k^2 + 2 \sigma(1-r)\} + 
2 \lambda^3\{ 1+ \sigma\} + \lambda^4\right]
\ee
with the abbreviation $\Omega_k:= \Omega_0- a k^2$.
To determine the instability threshold $r=r_c$ 
one inserts $\lambda=i \omega_c$ and obtains, by separating the real and 
imaginary part, the critical frequency 
\be{he}
\omega_c^2 = \frac{\Omega_{q_c}^2 + (1-r_c)(1+\sigma)\sigma}{1+\sigma}
\ee
and the threshold
\be{hf}
r_c=1+\left(\frac{\Omega_{q_c=0}}{1+\sigma}\right)^2 = 1+\omega_c^2 \quad .
\ee
The instability occurs at $q_c=0$ since the threshold (\ref{hf})
is minimal for that wavenumber. The right and left eigenvectors
(cf.~eqs.(\ref{bdc}) and (\ref{bf})) corresponding to the eigenvalue
$i \omega_c:=i\Omega_0/(1+\sigma)$ can be read off from the 
matrix as
\be{hg}
\underline{u}_c = \left( i, 1, i(1-i \omega_c),
1- i \omega_c, 0\right)^T, \quad 
\underline{v}_c^*=(1,i,\frac{\sigma}{1+i \omega_c},
i\frac{\sigma}{1+i \omega_c},0) \quad .
\ee

Introducing the deviation from the threshold by $r = r_c + \varepsilon^2 
\delta$, the perturbation ${\cal L}^{(2)}$ has only two nonvanishing matrix
elements and eq.(\ref{bn}) yields
\be{hi}
\eta= \delta \frac{(\ul{v}_c^*)_3 (\ul{u}_c)_1 + (\ul{v}_c^*)_4 
(\ul{u}_c)_2}{
( \underline{v}_c | \underline{u}_c )} = \delta
\frac{\sigma}{1+\sigma+ i \omega_c(1-\sigma)} \quad .
\ee

Keeping in mind, that $\ul{\ul{L}}(k)$ depends solely on
$k^2$, it is obvious that the convective velocity $v$ vanishes. The second
derivative, i.e.~the diffusion constant, is easily expressed in terms of
the eigenvectors (\ref{hg}) by tracing back to the usual Schr\"odinger
perturbation expansion
\be{hj}
D = -a \frac{(\ul{v}_c^*)_1 (\ul{u}_c)_2 - (\ul{v}_c^*)_2 
(\ul{u}_c)_1}{( \underline{v}_c|
\underline{u}_c )} =i a \frac{1+i \omega_c}{1+\sigma + 
i \omega_c(1-\sigma)} \quad .
\ee

Because of eqs.(\ref{hk}) and (\ref{hg})
$C^{(00)}\{\underline{u}_c,\underline{u}_c\}$ vanishes identically,
and $C^{(00)}\{\underline{u}_c,\underline{u}_c^*\}$ has only one nonvanishing
component
\be{hja}
C^{(00)}\{\underline{u}_c,\underline{u}_c^*\}= \left(0,0,0,0,
2 \right)^T \quad .
\ee
Then by virtue of eqs.(\ref{bmc}) and (\ref{bmd}) we have
$\Gb=0$ and
\be{hl}
\Ga=-\frac{1}{\ul{\ul{L}}(0)} C^{(00)}\{\ul{u}_c,
\ul{u}_c^*\}=\frac{2}{b} \left( 0, 0, 0, 0, 1\right)^T \quad ,
\ee
so that eq.(\ref{bmb}) reads
\be{hm}
\ul{\Gamma}= 2 C^{(00)}\{\ul{u}_c,2\Ga\}= \frac{4}{b}
\left(
0, 0, -(\ul{u}_c)_1, -(\ul{u}_c)_2, 0 \right)^T \quad .
\ee
Since $\ul{\Delta}=0$ because of the absence of cubic nonlinearities
we obtain finally for the nonlinear coefficient (\ref{bma})
\be{hn}
r= \frac{4}{b} \frac{-(\ul{v}_c^*)_3 (\ul{u}_c)_1 - 
(\ul{v}_c^*)_4 (\ul{u}_c)_2}
{( \underline{v}_c|\underline{u}_c )}
= - \frac{4}{b} \frac{\sigma}{1+\sigma + 
i \omega_c(1-\sigma)} \quad .
\ee

The formal simplicity of this example suggests, that 
the explicit evaluation is not at all a difficult task,
even in cases which are usually believed to involve lengthy calculations
like the Rayleigh--B\'enard  problem.
}

\end{document}